\documentclass[twocolumn,prb,showpacs,preprintnumbers,superscriptaddress,amsmath,amssymb]{revtex4}

\usepackage{graphicx}% Include figure files
\usepackage{dcolumn}% Align table columns on decimal point
\usepackage{bm}% bold math
\usepackage{color,pstcol}

\begin{document}

%\preprint{Draft ver. 0.95, 2003.12.22}

\title{Fluctuations, line tensions, and correlation times of nanoscale islands on surfaces}

\author{F. Szalma}
\email[Corresponding author: ]{szalmaf@physics.umd.edu}
\affiliation{
Department of Physics, University of Maryland, College Park MD 20742-4111}
\affiliation{
Institute of Theoretical Physics,
Szeged University, H-6720 Szeged, Hungary}
\author{Hailu Gebremariam}
\author{T.L. Einstein}
\email{einstein@umd.edu}
\homepage{http://www2.physics.umd.edu/~einstein}
\affiliation{
Department of Physics, University of Maryland, College Park MD 20742-4111}
\date{\today}% It is always \today, today,

\begin{abstract}
We analyze in detail the fluctuations and correlations of the (spatial)
Fourier modes of nano-scale single-layer islands on (111) fcc crystal surfaces.
We analytically show that the Fourier modes of the fluctuations couple due to the
anisotropy of the crystal, changing the power spectrum of the fluctuations, and that the
actual eigenmodes of the fluctuations are the appropriate linear combinations of the
Fourier modes. Using kinetic Monte Carlo simulations with bond-counting parameters
that best match realistic energy barriers for hopping rates, we
deduce absolute line tensions as a function of azimuthal orientation from the analyses of the fluctuation of each
individual mode.
The autocorrelation functions of these
modes give the scaling of the correlation times with wavelength,
providing us with the rate-limiting kinetics driving the fluctuations, here step-edge
diffusion.
The results for the energetic parameters are in reasonable
agreement with available experimental data for Pb(111) surfaces, and we compare
the correlation times of island-edge fluctuations
to relaxation times of quenched Pb crystallites.  
\end{abstract}

\maketitle

\section{Introduction}

Nanoscale islands consisting of $10^2$ -- $10^5$ atoms have
captured great interest over recent years for a variety of
reasons. From a practical standpoint, they provide a precursor to
the formation of quantum dots, which, if assembled in a controlled
way, can serve as the basic ingredients of nano-scale electronic
and mechanical devices. Many crystallites or nanomounds are best
viewed as ``wedding-cake"-like stacks of such islands.
\cite{pimpvillainbook} They are
the intermediary between a flat surface and a small
three-dimensional structure. In contrast to steps, which require
vicinal surfaces\cite{jeongwilliams99} that often must be well characterized over
mesoscopic regions, islands can be studied in smaller-scale
regions that are flat only locally.

Of particular interest to us are the shape and the fluctuations of
the perimeter of these islands.  The shape provides information
about the line tension or step free energy per length, from which
one can compute the step stiffness that describes the ``inertial"
properties of steps. The ``dipole" mode of these fluctuations are
known to underlie the diffusion of such islands, a concept now used routinely in 
simulations.\cite{Mueller99,Morg02}  However,
shorter-wavelength modes are also of great interest, since they
can be correlated with similar fluctuations of steps and provide
a way to assess, again, the stiffness of the step and also the
kinetic or atomistic diffusion coefficient associated with the
mechanism that dominates the atomistic processes underlying the
fluctuations.  Until recently, attention was limited to
structures for which crystal anisotropy could seemingly be ignored.

Here we pay particular attention to the role of the inevitable
anisotropy of crystal surfaces, which around room temperature or
even above it is typically sufficiently strong that it should
apparently be taken into account in order to correctly
characterize the morphology of the various (near) equilibrium
structures appearing on surfaces and their dynamics.  In this
paper we focus on the line tension and stiffness and their
orientation dependence; we give an analytic method to calculating
these physical parameters from the fluctuation of nanoscale
islands.

The little experimental data on such systems involve runs of
worrisome duration or use probes that provide scanned rather than
instantaneous images.  To generate fully-characterized data, we
turned to kinetic Monte Carlo (KMC) simulations to mimic the
equilibrium fluctuations of islands. These simulations are the
input of our analytic theory which, starting from the excess free
energy corresponding to the capillary wave fluctuations of the 
island edge, provides
the {\sl eigenmodes} of these fluctuations. Since the 2D Wulff
plot relating the equilibrium island shape and the line tension
in the azimuthal directions on the surface provides only {\it
relative} line tensions for various orientations, a key problem
is always the determination of the chemical potential $\lambda$
of the island edge, which then produces an {\it absolute}
relation. This potential can be determined with surprisingly good
($\sim10$\%) accuracy from the spectrum of the modes of the system. We
compare these eigenmodes and the simple Fourier modes of the
fluctuations and reach the (perhaps) surprising conclusion that the
anisotropy only affects the longer wavelength modes.

Another aim of the paper is to examine the correlation of the
fluctuations of the Fourier modes and thereby to find the
rate-limiting process driving the fluctuations in a fairly
realistic model. For our KMC simulations we sought a system for
which one could compute hop rates with good accuracy and for
which there was quantitative experimental data with which to compare.
Accordingly, we have chosen Pb(111) so as to be able to compare
with intriguing recent experiments by Th\"urmer {\it et
al.}\cite{thuermeretal01} This analysis gives the scaling of the
correlation time with the wavelength, that is the dynamic 
exponent $z$, and provides us with
characteristic times measured not only in MC steps, but in {\sl
real} time. Thus, we can compare directly with experiments and
extrapolate to different structures from the simple one
considered here.

Utilizing direct surface imaging techniques, especially scanning 
tunneling microscopy (STM),
several attempts have been made to measure and calculate step
energies. From a theoretical viewpoint the various methods that
used the experimental data for calculations can be broken down
into two main groups. The first is based on a lattice model
which relates the island shape (radius and curvature) to the temperature
dependence of the free energy and stiffness of the Ising model 
in the low-temperature expansion, usually in high
symmetry directions. By fitting the functional shape of the free
energy with varying temperature on the experimental data
determined by the equilibrium island shape \cite{schulzeetal99}
gives the Ising kink energy, which in turn
provides the step energies and stiffnesses. However, limitations of the Ising 
model to describe surface structure have recently been noted.\cite{dieluweit02}

The other method is based on a step continuum model which
makes use of stochastic differential equations to describe the
fluctuations of straight steps \cite{barteltetal93} or island
edges\cite{kod-khare,kod-pet} viewed as nearly circular closed-loop steps. Thus, the
initial calculations for island fluctuations assumed
isotropy:\cite{khareein96} the power spectrum of the Fourier
modes of the step fluctuations were calculated and adapted with
appropriate modifications to nearly circular island
shapes.\cite{schlosseretal99} If the anisotropy turns out to be
strong, it cannot be handled as a perturbation; a complete
anisotropic calculation without any such assumptions becomes
necessary.

\begin{figure}
\includegraphics[width=.9\columnwidth]{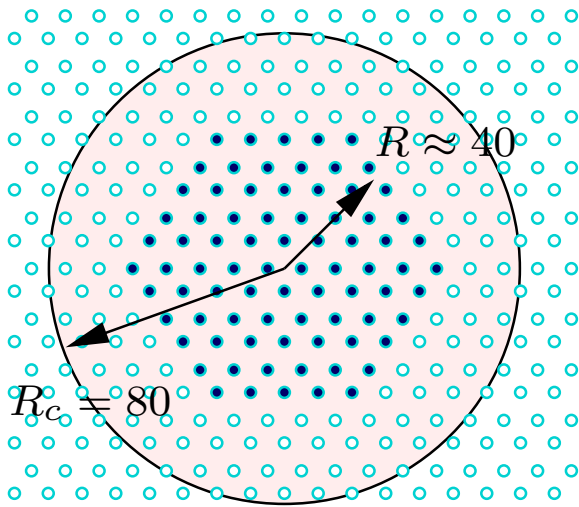}
%\resizebox{.8\columnwidth}{!}{\input{mcgeometry.pstex_t}}
\caption{\label{fig:mcgeometry} (Color online) Geometry of the MC simulation.
The approximate mean radius $R$ of the island and the radius
$R_c$ of the container are illustrated.}
\end{figure}

This challenge was recently taken up by Khare {\it et al.},
\cite{khareetal03} who give an approximate form for the free energy
functional and calculate the chemical potential integrating all
the Fourier modes in the system by using the generalized
equipartition theorem where the modes are buried in a sum.
However, these modes are coupled, so any one mode 
 missing (e.g.\ due to lack of experimental
resolution) in the sum can contribute to a
deviation from the precise value of the chemical potential by
itself and through its coupling to the other modes as well. In
contrast, our approach of analyzing individual modes gives
more insight into the extent to which this coupling should be taken into
account, and provides the chemical potential in a (mathematically)
controlled way.

The autocorrelation function of fluctuations of step edges and
correlation times have been analyzed theoretically in Fourier space
based on Langevin formalism,\cite{khareein96,khare98}
and experimentally in the context of straight-step fluctuations on
Si(111)\cite{bartelt96} and Si(001)\cite{barteltetal93} surfaces
for relatively long wavelengths. The rate-limiting kinetics
driving these fluctuations are determined by the dynamic
exponent, which also sets the universality class to which the system
belongs.\cite{barastanley95} (The roughness exponent is believed
to be
$1/2$ in our cases.) The correlation times are theoretically
identical to the relaxation times (or, in some cases, decay times) of surface
features,\cite{pimpvillainetal93,baleszangwill90} such as decay and near-equilibrium
build-up  of bulges (of either sign) along the step edge that also have (wave)length $L$. Three-dimensional
features like mesoscopic (or smaller) wires on surfaces as well as the
surface corrugations in
earlier studies by Mullins,\cite{mullins5759,mullinsbook} are typically
described by 1+1 dimensional models but may involve different, more
complicated mechanisms driving their
fluctuation or decay. We will compare and discuss these various relaxation times
in the paper.

The paper is organized as follows: In the next section we give an
analytic solution to the decoupling of the Fourier modes of the
system into the actual eigenmodes and recalculate the free energy
functional of the edge fluctuations. The results and conclusions can be
understood without the reader's going through this algebra; only the result 
expressed in Eq.\ (\ref{eq:eigmodes}) is used later. In Sec.\ III we introduce the KMC
simulation and in Sec.\ IV use its results to calculate the chemical
potential and line tension. In Sec.\ V we calculate the correlation
functions of the Fourier modes and deduce the scaling of the
correlation time with length, the dynamic exponent $z$. We compare 
with available experimental data for Pb(111). Sec.\ VI 
concludes the paper.

\section{Fourier modes, eigenmodes}

\begin{figure}
\includegraphics[width=.9\columnwidth]{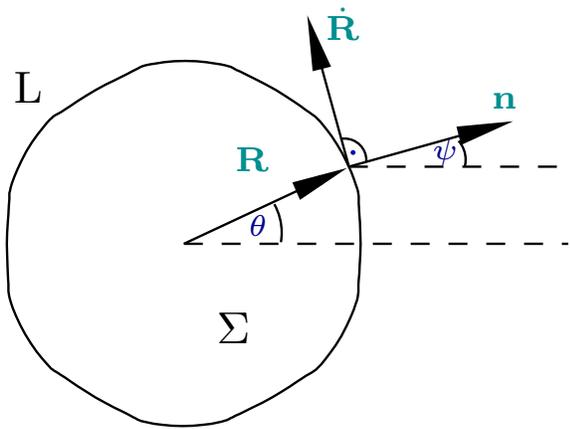}
%\resizebox{\columnwidth}{!}{\input{eqshape.pstex_t}}
\caption{\label{fig:eqshape} (Color online) Schematic showing variables used to analyze
equilibrium island shape}
\end{figure}

The relationship between the equilibrium crystal shape and the
surface tension or, in our 2D case, between the equilibrium island
shape and the line tension of its edge can be established by the
minimization of the free energy functional of the
island edge. The orientation-dependent line tension $\beta
({\rm\bf n})$ is defined as the work per unit length necessary to
create the $ds$ line element with normal ${\mathbf n}$ to the
perimeter. The free energy is the integral of this work along the
whole perimeter. The equilibrium island shape at a constant temperature $T$,
number of particles $N$, and area $\Sigma$,  is determined by the 
minimization of the free energy functional with respect to the
shape with the constraint that the island area is constant, typically using 
the method of Lagrange multipliers: \cite{landau}
\begin{eqnarray}
F[R,\dot{R},\theta]&&=\oint\limits_{L_{eq}}\beta ({\rm\bf
n})ds-\lambda \int \limits_{\Sigma }d\sigma = \nonumber \\
=&&\!\!\!\!\!\!\!\int\limits_0^{2\pi}\!\!\beta (\psi (\theta ))
\left(R^{2}+\dot{R}^{2}\right)^{1/2}\!\!d\theta
-\lambda\!\!\int\limits_0^{2\pi} \frac{R^{2}}{2}d\theta.
\label{eq:FreeenR}
\end{eqnarray}
Here the second line is in polar coordinates with $\theta$
the polar angle and $R(\theta)$ the radius of the equilibrium
shape. The dot denotes the differentiation with respect to the
angle, $\lambda$ is the Lagrange multiplier (which actually turns
out to be the chemical potential), and $\psi$ is the angle which
characterizes the vector normal to the shape (see 
Fig.\ \ref{fig:eqshape}). $ds$ and $d\sigma$
are the line element and surface element, respectively. Formally
minimizing the $F=F[R,\dot{R},\theta]$ functional, the
Euler-Lagrange equation gives a relation between the equilibrium
island shape $R(\theta)$ and the orientation dependent line
tension, $\beta(\psi)$, and between the two angles involved:
$\psi$ which depends on the polar angle and the equilibrium shape.
\cite{khareetal03,kodambakaetal02}
\begin{equation} \frac{\delta
F}{\delta R} =0\Longrightarrow
\begin{array}{c}
\beta(\psi)={\lambda}\frac{R^2}{\left(R^2+\dot{R}^2\right)^{1/2}}\\
\psi=\theta-\arctan\frac{\dot{R}}{R} \label{eq:betapsi}
\end{array}
\end{equation}
\noindent However, in this procedure $\lambda$ is a prefactor and
cannot be determined, leaving the relation relative. Eq.
(\ref{eq:betapsi}) is the seminal Wulff construction in polar
coordinates.

\begin{figure}
\includegraphics[width=.9\columnwidth]{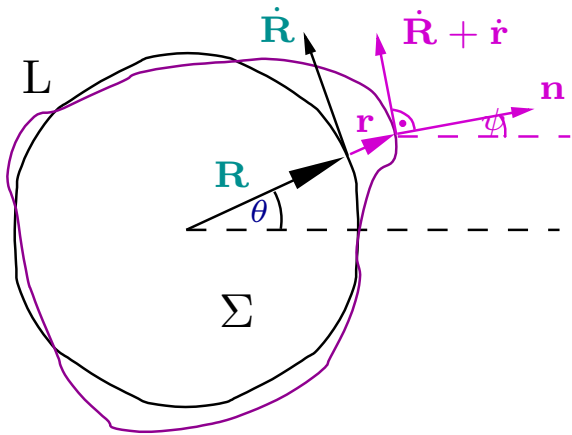}
%\resizebox{\columnwidth}{!}{\input{instantshape.pstex_t}}
\caption{\label{fig:instantshape} (Color online) Extension of Fig.~2, showing instantaneous island shape, for analysis of fluctuations. For a particular azimuthal
direction, $\theta$, the deviation from the equilibrium island shape, $r$, its derivative 
with respect to $\theta$, $\dot{r}$, the unit vector normal to the instantaneous shape, 
$\mathbf{n}$, and the corresponding angle, $\psi$, are all time dependent.}
\end{figure}

In order to determine the chemical potential, the thermal
fluctuations of the island edge can be utilized. In this case the
free energy of the island changes as its shape changes due to the
fluctuations, and the free energy is certainly not at its
minimum but depends on the island's instantaneous shape. Then
the free energy of this instantaneous shape is the integral over
the line elements of the shape with their corresponding line
tension, which changes with time as the orientation of the shape
element changes:
\begin{eqnarray}
F[r,\dot{r},\theta;t]&&=\oint\limits_{L}\beta ({\rm\bf n})ds= \nonumber \\
=&&\!\!\!\!\!\int\limits_0^{2\pi}
\!\!\!\beta (\psi (\theta ))\left((R+r)^{2}+(\dot{R}+\dot{r})^{2}\right)^{1/2}\!\!\!d\theta
\end{eqnarray}
Here $r$ and $\dot{r}$ are time dependent and describe the deviation of the 
instantaneous shape from the equilibrium one as shown in Fig.\ 
\ref{fig:instantshape}. The angle $\psi$ is also time-dependent
since now it depends not only on $R$ and $\dot{R}$ as in Eq.\ 
(\ref{eq:betapsi}), but also on $r$ and $\dot{r}$.
 Considering only small deformations from the equilibrium shape
(as it is usually assumed in the capillary wave theory) and also small slope
deviations from the equilibrium slope $\dot{R}$, so that $r,\dot{r}\ll R$, the Taylor expansion
(both in $\beta$ and in the square root) in these small parameters leads to the functional
\begin{equation}
F[r,\dot{r},\theta;t]=\lambda\int\limits_0^{2\pi}\frac{1}{2}
\frac{\left(\dot{R}r-R\dot{r}\right)^{2}}{R^2+2\dot{R}^2-R\ddot{R}}d\theta.
\end{equation}
This functional contains 3 quadratic terms $A(\theta)r^2$,
$Q(\theta)r\dot{r}$ and $B(\theta)\dot{r}^2$. The cross term $Q$
drops out after taking the ensemble average; both the two other
terms are determined by properties of the equilibrium island
shape:
\begin{eqnarray}
A(\theta)=\frac{1}{2}\frac{\dot{R}^{2}}{R^2+2\dot{R}^2-R\ddot{R}}\\
B(\theta)=\frac{1}{2}\frac{R^{2}}{R^2+2\dot{R}^2-R\ddot{R}} \; ,
\end{eqnarray}
and provide the weightings of the fluctuations of the deformations characterized 
by $r^2$ and $\dot{r}^2$, respectively. These deformations at the microscopic level
are due to the thermal movement of adatoms surrounding the island constantly 
attaching to its edge and coming off from it.

To diagonalize the free energy one rewrites the integrand in Fourier form
\begin{equation}
\hspace{-0.1mm}F[\{r_n\};t]=2\pi\lambda\sum_{{m},{n}}\left(A_{m-n}+mnB_{m-n}\right)r_{
{n}}(t)r_{{m}}^*(t)\; ,
\label{eq:FAB}
\end{equation}
where $r_k=\int_0^{2\pi}r(\theta)\exp[ik\theta]d\theta$ and similarly 
for $A_k$ and $B_k$. The Fourier modes
are coupled due to the anisotropy, which is contained in $A$ and $B$ as we 
shall see shortly. Here $n=0$ is the
expansion-contraction mode; $n=1$, which we called the dipole mode in the 
Introduction, is related to the Brownian,
diffusive motion of the island; $n=2$ is a quadrupolar
distortion, i.e.\  an elongated shape with two maxima and two
minima in perpendicular directions; and so on.
The Fourier components have hermitian properties since $A(\theta)$
and $B(\theta)$, the factors associated with the equilibrium
island shape, are real functions; hence, $A_{-i}=A_i^*$,
$B_{-i}=B_i^*$, and $r_{-i}=r_i^*$. 

The free energy of
Eq.~(\ref{eq:FAB}) can readily be cast into matrix form:
\begin{equation}
F[\mathbf{r};t]=2\pi\lambda
\mathbf{r}^\dagger\left(\mathbf{A}+\mathbf{MBN}\right)\mathbf{r}
\; ,
 \label{eq:rundiag}
\end{equation}
where $\mathbf{r}$ is a vector containing the Fourier components
of the instantaneous island shape, $\mathbf{A}$ and $\mathbf{B}$
are hermitian matrices, $[\mathbf{A}]_{m,n}=A_{m-n}$,
$[\mathbf{B}]_{m,n}=B_{m-n}$, and $\mathbf{M}=\mathbf{N}$ are
diagonal matrices with the wavenumbers along the diagonal. 

As in practice there are only a finite number of atoms on the edge of 
the island, we discretize the problem. If the number of atoms on the edge
is $2N$, there are as many modes in the system; as we will see in Sec.\ 
IV, to analyze the right number of modes is crucial to the problem.
Now, if $\mathbf{r}$ contains
the $r_k$ Fourier components from $-N+1$ through $N$, the Fourier transform is 
discrete and $r_k=\sum_{j=-N+1}^Nr_j^\theta\exp[i\,k\,j\pi/N]$, where $r_j^\theta$ 
is the deviation from the equilibrium shape in the $\theta=j\pi/N$ direction. 

The $A_k$ 
and the $B_k$ can be obtained similarly; and $\mathbf{A}$
and $\mathbf{B}$ are finite cyclic hermitian matrices, meaning that their
diagonal elements are the same. They also reflect the symmetry of
the equilibrium shape as e.g.\ in our case due to the six-fold
symmetry the principal diagonal is filled with $A_0$, the 6th to
the right with $A_{-6}$, etc. As $\mathbf{M}$ and $\mathbf{N}$ are
the same diagonal matrices, the $\mathbf{MBN}$ product keeps the
hermitian property.

In the isotropic case (when the equilibrium shape is circular),
$A(\theta)=0$ and $B(\theta)=1/2$ for all $\theta$. After the
Fourier transformation this gives $\mathbf{A}=0$ (zero matrix) and
$\mathbf{B}=(1/2) \openone$ (diagonal matrix). The anisotropy
comes into play when the equilibrium shape is not circular, so that
$A(\theta)$ and $B(\theta)$ are not constants and their higher
order Fourier components fill the (off-)diagonals. These
off-diagonals couple the Fourier modes.

Due to hermiticity the above matrix form is diagonalizable
\begin{equation}
F[\{h_n\};t]=2\pi\lambda\sum_{{n}}\Lambda_nh_{{n}} h_{{n}}^* \; ,
\label{eq:hdiag}
\end{equation}
and the eigenvalues $\Lambda_n$ of the $\mathbf{A}+\mathbf{MBN}$
matrix are all real. As we see shortly (in Eq.\ (\ref{eq:eigmodes})) 
these eigenvalues are related to the strengths of the $h_{{n}}$ 
{{eigenmodes}}, which at every time instant are just the
transforms of the $r_{{n}}(t)$ Fourier modes of the instantaneous
island shape. Again due to hermiticity, there is a unitary matrix 
$\mathbf{U}$ which transforms Eq.\ (\ref{eq:rundiag}) into Eq.\ 
(\ref{eq:hdiag}) and gives the 
linear relationship between $\mathbf{r}$ and $\mathbf{h}$: 
$\mathbf{r}=\mathbf{Uh}$, where the vector $\mathbf{h}$ contains the $h_n$ 
as its elements.

This decomposition of the free energy into eigenmodes in
Eq.~(\ref{eq:hdiag}) facilitates the calculation of the Lagrange
multiplier ${\lambda}$. In equilibrium, according to the
equipartition theorem, the ensemble average of each mode, 
representing a degree of freedom, must
have the same Boltzmann energy:
\begin{equation}
2\pi{{\lambda}}\overbrace{\Lambda_n\langle\left|h_n\right|^2
\rangle}^{E_n}=\frac{1}{2}k_BT.
\label{eq:eigmodes}
\end{equation}
$\Lambda_n$ and $\langle|h_n|^2\rangle$ can be determined from the equilibrium
island shape and the fluctuating island perimeter, respectively. Here $E_n$ must be a 
constant in $n$, the modes, as the temperature and the chemical potential, $\lambda$, 
are fixed macroscopic parameters of the island. From this equation one can determine the 
same $\lambda$, in principle, from any mode. Thus, either experimentally observing island 
fluctuations or using Monte Carlo simulations one can determine $E_n$, which in turn 
provides $\lambda$. This $\lambda$ was the missing parameter to determine {\sl absolute} 
line tensions, and plugging it back into Eq.\ (\ref{eq:betapsi}), we get the line tension in all azimuthal 
directions.

\section{Kinetic Monte Carlo}
The scarcity of extensive experimental data leads us to use Monte Carlo methods to
simulate the behavior of the system. 
However, use of numerical rather than experimental data for testing of formal ideas has many advantages in any case.  Most obviously, in numerical experiments one can obtain far greater control, with no worries about anomalous behavior due to unsuspected stray contaminants.  Typically one can generate much more data.  In the present experiment, we do not need to worry about the scan rate of the probe; our lattice configurations are instantaneous snapshots.
Another advantage of computer simulations will
also become clear in the next section: it allows us to analyze correlation times.

Since our original motivation was to simulate the relaxation of a
Pb crystallite with a (111) facet, we place a nanoscale island on
a triangular lattice. We surround it by a non-permeable circular
container of radius $R_c$ to let the system reach its
thermodynamic equilibrium, in order to measure its equilibrium
fluctuations.\cite{sethna,grabow} Thus, this geometry corresponds
to an island placed on top of a facet of a crystallite (with an
infinite Ehrlich-Schwoebel barrier).  Note that by adjusting the
permeability one can tune the overall decay rate of the island,
which in this paper we fix at zero.

\begin{table}
%\begin{ruledtabular}
\begin{tabular}{l||r|r|r}
Process & \shortstack{Energy\\ (meV)} & 
          \shortstack{Energy\\ (K)} & 
          \shortstack{Break-three energy\\ (K)}\\
\hline
Surface diffusion &  70 &   812 &   812\\
Edge diffusion    & 237 &  2749 &  2319\\
Break 1 bond      & 192 &  2227 &  2319\\
Break 2 bonds     & 359 &  4164 &  3826\\
Break 3 bonds     & 467 &  5417 &  5333\\
Attachment            &     &       &   812\\
Out               &     &       & 70000\\
\end{tabular}
\caption[shrtab]{Tabulation of some of the energy barriers used in KMC
simulations of Pb(111).  The energies in columns 2 and 3 were
computed by M. Haftel using SEAM\cite{haftel} with glue 
potentials.\cite{ercol-Pb}  In the last column are the energies 
used in the simulations.  (The edge-diffusion energy barrier in the 
Break-three scheme is closer to the corner rounding barrier of SEAM than to the actual 
straight-edge diffusion barrier of SEAM (see text); the latter has a much lower 
barrier: 108 meV; unfortunately, the bond-counting method does not
distinguish between these two.)
\label{tb:enPb}}
%\end{ruledtabular}
\end{table}

Since the temperature of the systems of interest is low compared
to the energy barriers of adatomic hopping, we have chosen to use
the Bortz-Kalos-Lebovitz (BKL) continuous-time MC algorithm\cite{BKL} as it
is best suited to low temperature systems and as its
rejection-free method allows us to greatly improve the efficiency
of the simulations compared to traditional Metropolis algorithms.
The typical temperatures are around $T_c/4$ or less for the two-dimensional 
lattice gas of adatoms on the surface.
Using the $n$-fold way method to keep track of the available MC
moves, we could improve the efficiency even further.  Because of
the small number of energy barriers, the $n$-fold way approach
(5-fold) is superior to the binary tree implementation of the BKL
algorithm\cite{NewBark} in this case.

Since we are not interested in all the 
details of this surface in the simulations, but only try to 
capture the main mechanisms, we do not take into account the 
ABC stacking structure of the fcc lattice of Pb(111). Hence, the top layer 
constitutes a triangular lattice with perfect six-fold symmetry.
 Furthermore, we assume that adatoms can only hop to 
nearest neighbor sites, and that the energy 
barrier the adatom has to overcome is determined by the occupation of the 
8 sites surrounding, as nearest neighbors, the 2 sites involved in 
the hopping process.

The energy barriers for hopping rates are mainly based on the
semiempirical embedded atom method (SEAM)\cite{haftel} using
Ercolessi's glue potentials\cite{ercol-Pb} for the Pb(111)
surface to derive the bond-counting energy barriers\cite{NewBark} 
actually used in the simulations (see Table \ref{tb:enPb}). 
We use variants of simple bond-counting.  In what we term the break-three scheme,  we count only 
bonds that have been broken with the three sites ``behind'' the 
move. 
If we also include bond-breaking with the 2 sites to the left and the right of 
the move (denoted side sites), we have a break-five scheme. 
The 3 sites in front of the move do not affect the energy barrier 
in our simulations. The break-three and break-five schemes both satisfy 
detailed balance in a straightforward fashion. They both should give the same results 
for static parameters, since they are both nearest-neighbor schemes. Comparison runs
using this feature provided one test (among many) of our program. However, the 
kinetics obviously differ because the energy barriers tend to be higher for 
the break-five scheme, slowing the kinetics significantly. Since the energy 
barriers in the break-three case are closer to the calculated barriers, we choose 
to use this scheme in our simulations. 

In Table \ref{tb:enPb} we list a few energy barriers calculated by the 
above-mentioned SEAM and the corresponding break-three-scheme barriers. 
Surface diffusion is when an adatom has no other adatoms in the surrounding 8 sites
in SEAM calculations; in the break-three scheme this barrier also applies to all 
cases in which any of the side sites or the sites in front are occupied. This is the 
reason why the attachment barrier is the same as the surface diffusion barrier. Edge 
diffusion is the case in which a side site is occupied, as is its nearest neighbor 
``behind''. The energy barrier associated with this process is 237 meV whereas if there 
is a nearest in the front as well so that the adatom rolls along  three others on one side 
the third in the front seems to assist the hop a great deal (at least according to the SEAM 
data) as the barrier is 108meV. The break-three scheme has the same barrier for these two 
processes and also for any other in which only one bond is broken. The break-two-bonds barrier 
corresponds to a hop with 2 occupied sites in the back, Break-3-bonds is when all 3 sites 
are occupied in the back. The very high ``Out'' energy barrier assures that adatoms cannot escape 
the container; i.e., the permeability is zero.

The basic parameters of the surface investigated and the KMC simulations 
are the following: The nearest-neighbor spacing $a_1$ on the Pb(111) surface is
3.50\AA. The typical island diameter $2R$ is 40$a_1$ to 80$a_1$, while
the container diameter $R_c$ ranges from 12.5\% to 300\% larger
than the island. We examined temperatures 250K, 300K, 350K, and 400K.
In each MC snapshot of the island, we measure the island radius
from the instantaneous center of mass in $N$ ``equiangular"
directions where $N=360$ if not indicated otherwise. 

We start the simulations from a nearly circular shaped configuration 
and let it relax to equilibrium, starting the MC measurement of the 
fluctuation and shape after the longest wavelength mode has passed 
its correlation time. Especially at the lower temperatures, the typical
equilibration times are very long, consistent with other reports.\cite{szalma01, rahman03} Ensemble averages are taken 
from 100 to 3000 different runs starting from the same initial 
configuration, but with different random-number seeds. In each run, 
after equilibration, we get statistically independent fluctuations at 
time intervals again determined by the relaxation time of the longest wavelength 
mode. We take such independent ``snapshots'' of the islands 5 to 200
times in each run, so that we typically have 10,000 to 70,000 islands 
over which to average.

\section{Chemical potential, line tension, line stiffness}

As described in detail in Sec.\ II, the energetic parameters of
the island edge are determined by the island shape and its edge
fluctuation. The Wulff construction provides the relationship
between the relative line tension in the azimuthal directions on
the crystal surface and the equilibrium island shape, and the
information from the fluctuations
$\langle\left|h_n\right|^2\rangle$ of each mode in Eq.\ (\ref{eq:eigmodes})
gives the chemical potential $\lambda$ that makes the Wulff
construction absolute in Eq. (\ref{eq:betapsi}).

\begin{figure}[t]
\includegraphics[width=.9\columnwidth]{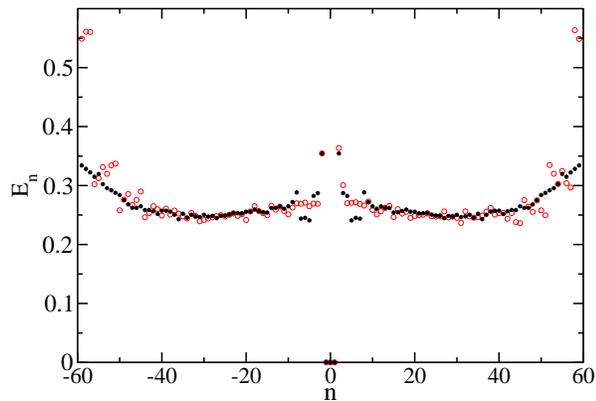}
\caption{(Color online) Eigenmodes (open squares) and Fourier modes (solid
circles) at $T$=250K, $R$=20$a_1$, $R_c$=40$a_1$. $E_n$ is measured
in atomic spacing units.} \label{fig:T400R80r20eigenFouriermodes}
\end{figure}

From our KMC simulations we determine $E_n$ of Eq.
(\ref{eq:eigmodes}); it is depicted in Fig.
\ref{fig:T400R80r20eigenFouriermodes} for $T$=250K and $R$=20$a_1$
island radius. Since the perimeter is about 120$a_1$, we use 120
points to describe the circumference out of the 360 available.

We calculate $E_n$ both using the transformation to the
eigenmodes taking into account the anisotropy, and also pretending the
islands were isotropic. In this latter case the $h_n=r_n$ are
simply the Fourier modes, and $\Lambda_n=(1/2)n^2$.

 The Fourier modes and the eigenmodes are nearly indistinguishable (cf.\
Fig.~\ref{fig:T400R80r20eigenFouriermodes}) {\it except} for long wavelengths
(small wavenumber $n$). The ``fluctuations'' in the Fourier modes for
longer wavelengths are the only signature of the anisotropy. This effect
is smoothed out by transforming to the eigenmodes. (Only the longest [$n$=2]
wavelength mode seems to stand out after the transformation. We are still
puzzled that the longest wavelength is so special and does
not couple to the shorter wavelength modes. We give a more detailed account
of the analysis of the coupling elsewhere.\cite{szalmaetal04}) The chemical
potential can be determined from either the Fourier modes or the eigenmodes
using their plateau regions (between $n=4$ and $n=40$); however, since
only the longest-wavelength modes (if any) are measurable due the poor temporal
resolution of present-day experimental apparatus (see the results for correlation times
in the next section), the transformed modes serve better for determining
chemical potentials. Finding $E_n$ thus from these intermediate
wavenumbers, gives
$E_n=$0.275$a_1^2$ and through Eq.\ (\ref{eq:eigmodes}) and Eq.\
(\ref{eq:betapsi})
$\beta=$34.1~meV/\AA\ for the line tension for the
high-symmetry direction. This value reasonably approximates the
experimentally obtained ones for Pb(111) at T=393~K:
$\beta_{1A}=27.9$~meV/\AA\ and $\beta_{1B}=26.5$~meV/\AA\ for A-
and B-type steps, respectively, considering the crude approximation of
bond-counting mentioned earlier.\cite{bombis-bonzel} In our
simulations the two directions corresponding to the two different types
of steps are intrinsically equivalent because we assume six-fold symmetry
as the available values for energy barriers that we use in
our KMC do not distinguish between the A- and B-directions, as
mentioned in Sec.\ III.

At higher temperatures the Fourier modes deviate less from the eigenmodes
for longer wavelengths, as expected since the equilibrium shape
is more nearly circular and less affected by the underlying anisotropy.

In earlier work\cite{khareetal03,kodambakaetal02} in which
experimental data were used as an
input of similar calculations, there is a sum over the modes, but
because those modes are buried in a sum in the generalized
equipartition theorem, one cannot see whether they are the modes
which satisfy, at least to a certain extent, the equipartition
theorem. In those experimental data the correlation times of the modes are not 
known and the effect of the finite temporal resolution of the experiment 
may also interfere with the fluctuations which should in principle be determined 
from ``snapshots'', i.e. fast scanned images---fast at least compared to the 
correlation times of the modes used in such calculations. We shall elaborate 
on this in the next section.

\begin{figure}[t]
\includegraphics[width=.9\columnwidth]{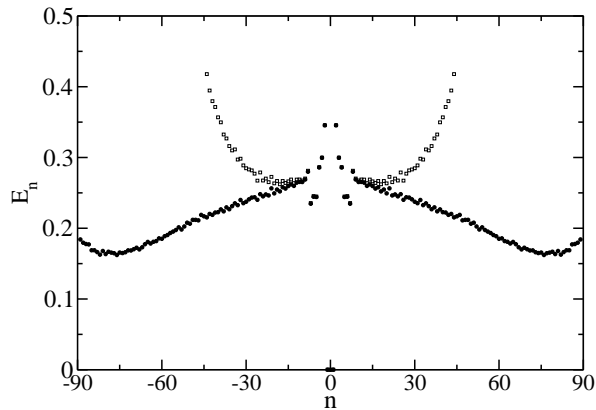}
\caption{Fourier modes for $N$=90 points (open squares) and
$N$=180 points  (solid circles) on the perimeter at $T$=250K, and 
$R$=20$a_1$, $R_c$=40$a_1$. $E_n$ is measured in atomic spacing
units.} \label{fig:T400R80r40Fouriermodesn180n360}
\end{figure}

For the same temperature we do the same measurement as above, but monitored $N=180$ 
points on the perimeter instead of $N=360$. The Fourier modes are depicted in Fig.\ 
\ref{fig:T400R80r40Fouriermodesn180n360}. There are
approximately 120 atoms on the perimeter, but since we cannot divide the 180
perimeter points into 120 equiangular ones to do a Fourier
transforms, we use $N$=90 or $N$=180 perimeter points as an approximation and observe
how the plateau changes from what we saw in Fig.\
\ref{fig:T400R80r20eigenFouriermodes}. The comparison of these two 
plots from MC simulations might help analyzing experimental
data with limited spatial resolution as well, as it shows how the Fourier modes
behave in case of undersampling ($N$=90) and oversampling
($N$=180). The undersampled modes give higher values for $E_n$
than expected for modes $|n|>25$, as if those modes took over the
energy of the modes that are missing from spectrum (namely $45<|n|\leq90$). 
In oversampling
there is not enough energy for all modes in the sampling, so they
go below the expected value of $E_n$. This is the simple reason for the 
peculiar shape of the two curves in Fig.\ 
\ref{fig:T400R80r40Fouriermodesn180n360} The value of $E_n$ can still be 
determined quite accurately by using the value at which the two curves start to separate at $|n|\simeq10$, 
providing basically the same result for $E_n$ as before.

\begin{figure}[b]
\includegraphics[width=.9\columnwidth]{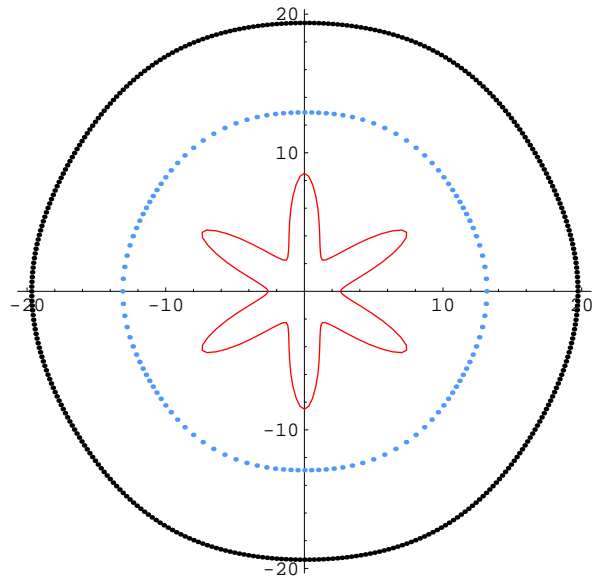}
\caption{(Color online) Polar plot of the equilibrium island
shape $R(\theta)$ (outer dots), the relative line tension 
$\beta(\psi)$ (inner dots) and the relative line stiffness 
$\tilde\beta(\psi)$ (the innermost curve)  in arbitrary units at $T=$350K, $R$=20$a_1$,
$R_c$=40$a_1$. (note the difference between $\theta$ and $\psi$)}
\label{fig:T400R80r20shapebeta}
\end{figure}

From the equilibrium island shape using the Wulff construction, we have 
determined the relative line tensions and stiffnesses in the azimuthal directions on the 
(111) surface (see Fig. \ref{fig:T400R80r20shapebeta}).
The equilibrium shape is more and more ``faceted'' in the six main directions as we can expect, 
the stiffness is about 3 times bigger in the direction of the ``facet'' than in the direction of 
the corner for $T$=350~K, whereas this factor is about 20 for $T$=250~K, so that it
spikes out  much more, but still has only a smaller effect on the spectrum as shown before. 

\section{Autocorrelations, correlation times, kinetics}

Inspecting the autocorrelation functions in Fourier space, we find that the
longest wavelengths have surprisingly long correlation times (in
CPU time). Thus, for our fairly large
(at least for the computer) systems, it is hard to reach full equilibration needed to make the desired MC measurements. Most surprising is that the relaxation time of the longest-wavelength modes
is 10 to 100 times longer than the relaxation time of the islands
to their equilibrium shape. Hence, estimating the
thermalization time from just the shape relaxation may be very
misleading, possibly giving problematic results not characteristic
of equilibrium. Such behavior may include illusory strong-mode coupling,
or stronger fluctuations in autocorrelation functions even in case of good 
statistics.

The temporal correlations can be characterized by
\begin{equation}
G(t)=\left\langle
\left[r(t_{0})-r(t_{0}+t)\right]^{2}\right\rangle \propto
t^{2\beta}.
\end{equation}
Since we measure correlations in equilibrium, $t_0$ must be greater
than the thermalization time of the system. Here $r(t)$ is the
fluctuation from the equilibrium shape, as before, and depends on
the angle, $\theta$, and time. The growth exponent\cite{barastanley95}$\beta$  
characterizes the temporal behavior of the fluctuations. 
The average is taken over angles and an ensemble as well.
%The angular averaging is strictly valid only for the isotropic
%case, but the results of the previous section imply that it is
%adequate in our case.(WHAT ABOUT IN KHARE'S?)

The typical behavior of the correlation function is that the
exponent $\beta$ remains at 1 for very short times,\cite{szalma01}
typical of the ballistic behavior of diffusion 
at very short times, and then
crosses over to a value which characterizes the rate-limiting
kinetics driving the fluctuations of the island edge; eventually
it crosses over to zero as the correlation function saturates due to 
the finite size of the system.

Pure rate-limiting kinetics have been thoroughly investigated.
\cite{BGEW92,BEW94,khareein96,khare98,blag-dux}
In these well-defined cases, $\beta$ can take the values 1/4 for 
attachment-detachment kinetics, 1/6 for surface diffusion, 
and 1/8 for step-edge diffusion, where the last mechanism gives a 
very ``slow'' dynamics. There can be crossover regimes between these
pure cases, leading to values of $\beta$ between the quoted values,
and certain geometries can also effect the value of $\beta$. One
should also see crossovers as length scales
vary.\cite{khareein96,khare98,blag-dux}

To investigate the length-scale dependence of the correlation
function, it is more appropriate to use the correlation function
in Fourier space:
\begin{eqnarray}
G_{n}(t)=&&\!\!\!\left\langle \left|r_{n}(t_{0})-r_{n}(t_{0}+t)\right|^{2}\right\rangle\\
=&&\!\!\!C_n\left(1-\exp\left(
-\left|t\right|/\tau_n\right)\right)\; ,
\end{eqnarray}
where the $C_n$ are twice the amplitudes of the fluctuations of
the modes analyzed in the previous section, and the $\tau_n$
are their correlation times. The wavenumber dependence of
$\tau_n$ is known to have an intimate relationship with the 
exponent in $G(t)$, namely
\begin{equation}
\tau_n \sim n^{-z},
\end{equation}
where $z$ is the dynamic exponent, and the scaling relationship between 
$z$ and $\beta$ here is $z=(1/2)/\beta$.\cite{barastanley95} The correlation time 
increases with increasing wavelength,
with the scaling exponent $z$. For larger exponent $z$, the
correlation times grow more rapidly, so that for longer
wavelengths the correlations and the dynamics in general 
can slow down very ``quickly''.

\begin{figure}[t]
\includegraphics[width=.9\columnwidth]{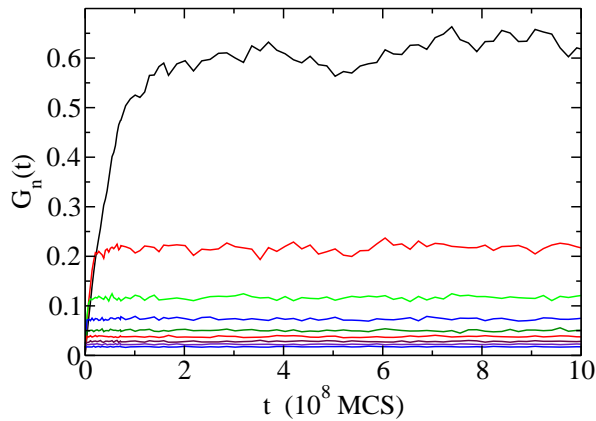}
\caption{(Color online) Correlation
function $G_n(t)$ of the Fourier modes for $n$=2,3,...,10 from top
to bottom. $T=$400K, $R$=20$a_1$, $R_c$=40$a_1$. } 
\label{fig:T400R80r20sincoscorrvst}
\end{figure}

Here we pay particular attention to the longest wavelength and its 
corresponding correlation time $\tau_2$, which makes the largest contribution 
to the fluctuations and relaxes the most slowly. From the wavelength dependence 
of the correlation time, we also determine the dynamic exponent and 
the rate limiting kinetics.

In the KMC simulations for $T$=400K and $R$=20$a_1$, the
longest-wavelength mode, $n$=2, is 120$a_1$ or 420\AA\ long. From Fig.
\ref{fig:T400R80r20sincoscorrvst}, its relaxation time is
$\tau_2=5.5\times10^7$ MCS (Monte Carlo steps). To give a crude
estimate for $\tau_2$ in real time, we consider the hopping rate
\begin{equation}
\nu=\nu_D\exp[-\beta E_b]
\label{eq:active}
\end{equation}
to be the product of the attempt frequency, which we identify with
the Debye frequency of Pb: $\nu_D=1.83\times10^{12}$
Hz,\cite{ashcroftmermin76} and the Boltzmann factor of the energy
barrier of a particular hop. Hence, a MCS in this Monte Carlo 
simulation is equivalent to a $1/\nu_D$ time increment in {\sl real} 
time; thus, the relaxation time in this particular case is $\tau_2=0.030$ msec.

As expected, these correlation times change dramatically with
temperature as the underlying physical phenomena are
activated. For $R$=20$a_1$ at T=350~K, $\tau_2=2.0\times10^{8}$
MCS or 0.11~msec, which means 4 times longer relaxation compared
to 400~K, while for 300~K $\tau_2=7.1\times 10^8$ MCS or 0.39~msec,
which represents slowing by another factor of 4.

We have to mention here that we did two sets of kMC simulations. 
The simulations described in this paper satisfy detailed balance and sample 
the canonical distribution, while in the other set we choose the energy barriers to be 
those calculated by SEAM, which explicitely violate detailed balance and energy conservation.
Interestingly, the latter simulations tend to give closer agreement with experimentally measured 
correlation time data.
We do a comparison of the results of these two sets of simulations\cite{szalmaetal04} and upcoming 
experimental data\cite{doughertyszalma04} elsewhere. 

\begin{figure}[t]
\includegraphics[width=.9\columnwidth]{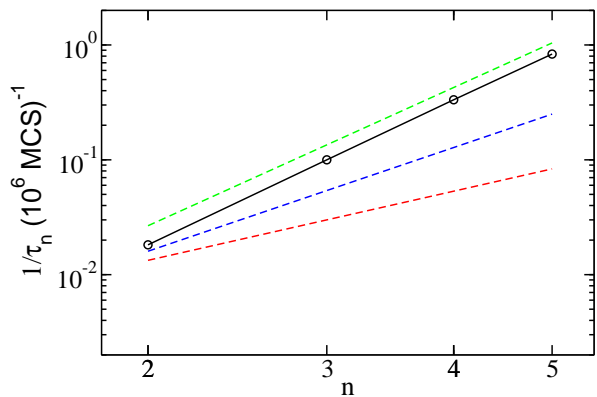}
\caption{(Color online) Correlation time
$\tau_n$ vs. wave number $n$ on a log-log scale. The MC data
(black circles) show  $z=4$ which corresponds to step-edge diffusion. The dashed lines
represent $z$=4, 3, and 2 (from top to bottom) dynamic exponents.
$T=$400K, $R$=20$a_1$, $R_c$=40$a_1$.} 
\label{fig:invtauvsq4008040}
\end{figure}

The scaling of the relaxation time with (wave)length can be seen in 
Fig.\ \ref{fig:invtauvsq4008040}. In the plotted wavenumber range, 
overall, $\tau_n$ basically behaves like $z=$4, suggesting that the
mechanism driving the fluctuations is step-edge 
diffusion which is in agreement with previous observations.\cite{kuipersetal95,spelleretal95} 

Comparison of these length scales and their corresponding
relaxation times with existing experimental observations might
give interesting physical insight. For example, in the experiment
by Th\"urmer {\it et al.},\cite{thuermeretal01} a small Pb
crystallite, whose top facet has a perimeter slightly larger than 1200 nm, 
relaxes at 383K to its
equilibrium (or at least metastable) shape in 1-2 days after
being quenched from a higher temperature. 
 Since the crystallite is 30 times larger than the
longest wavelengths in our KMC simulations, the relaxation time 
$\tau_{relax}$ of the longest wavelength mode of the perimeter of the topmost 
island on the crystallite is $8.1\times 10^5$ times  
longer if the rate-limiting kinetics is step-edge diffusion
(though, of course, attachment-detachment and terrace diffusion
could be present but NOT rate-limiting).  Specifically, the
value of $\tau_{relax}$ is 24.3~sec 
based on our KMC data at $T$=400K. From these data it seems that 
the island fluctuation is much faster than the decay of the 3D 
structure; thus there is no direct relationship between the fluctuation and the decay.

The above arguments lead to a general view of the evolution of
surface structures. For Pb in the temperature range 350 K--400 K, 
one observes the slow development and relaxation of fluctuations at 
the $\mu$m scale in experiments. Assuming that the rate-limiting
kinetics retain the same $z=$4 range for even longer
wavelengths, structures of 10 $\mu$m size --- step edges,
islands, etc.\ --- take days to years to change due to the large 
dynamic exponent, $z$, so in effect they look
frozen under laboratory conditions. This is the reason why
these monolayer structures do not show any large-scale changes 
while on a shorter scale they can be very active. The structural 
changes in the 3D crystallites are even slower, they are even more stable.

Lowering the temperature makes the length scales --- at which
evolution or relaxation can be observed --- exponentially
shorter, which is readily understandable if one looks at the
converse of the above arguments. The length scales with time like
$l\sim\tau^{-1/4}$ (for $z=4$) whereas $\tau$ scales like $1/\nu$ in 
Eq.~(\ref{eq:active}). Thus, the length scales with temperature like
$l\sim\exp{[E_b/4k_BT]}$. This basically means that given the 
temperature and the time scale of observation, one can calculate 
an effective length scale, $L_{\rm eff}$ on which the structures on a surface 
are in equilibrium with their surroundings and actively changing on the 
time scale of e.g.\ STM measurements. Having $L_{\rm eff}$ and the energy 
barrier $E_b$ at a certain temperature, one can also make at 
least rough estimates of the effective lengths at other temperatures using 
this scaling argument. This picture is certainly a result of simplification, since
there is a whole set of energy barriers in such a complex physical 
system as a crystal surface, and the various atomistic mechanisms 
governed by different barriers freeze out or get activated at different 
temperatures, depending on their corresponding energy barriers.\cite{Bogi} 

Comparison or extrapolation to other materials is possible if the energy-barrier 
set is similar to that of Pb(111). Then the Debye frequency sets 
the time scale while the energy barriers set the temperature scale, as one 
can readily deduce it from Eq.~(\ref{eq:active}). On the other hand,
if the energy-barrier set is completely different, as for example for Si,
\cite{bermondetal95} it gives rise to a different rate-limiting mechanism 
namely attachment-detachment for a wide range of temperatures, and such 
extrapolation is not possible, but a whole set of simulations should be 
done for the group of materials with this sort of barriers.

 \section{Conclusions}

In this paper we deduce energetic and kinetic parameters of a
particular metal surface below its roughening temperature. We use
kinetic Monte Carlo simulations to mimic the fluctuations of
large nanoscale islands on these smooth surfaces in order to
determine equilibrium island shapes, anisotropic line tensions in
the azimuthal directions of the surface, and the correlation
times of the Fourier modes of the fluctuations.

We derive an analytic expression for the chemical potential of
the island edge from the equilibrium island shape and the
associated capillary wave fluctuations around it. This chemical
potential sets the scale for the anisotropic line tension, [the
azimuthal dependence of] which is usually known only up to a
multiplicative constant.  To account for the anisotropy of the
line tension, this procedure contains a transformation from the
Fourier modes of the island edge fluctuations to the true
eigenmodes. However, detailed analysis of the Fourier and eigen
modes of the fluctuations reveals that the difference in their spectrum 
(Fig. \ref{fig:T400R80r20eigenFouriermodes}) is unexpectedly small.

The obtained line tensions---one of the most important physical
parameters of steps on surfaces---are in the correct range
compared to known experimental results, even in this simplistic
model, with its rather small set of hopping-energy barriers in the
KMC simulation.

We have analyzed the effect of spatial sampling, which shows
that the long wavelength modes are hardly affected by the undersampling
(oversampling)---too low (too high) resolution of imaging, which means that
there are fewer (more) sample points on the step edge in the image
than actual atoms in the experiment---
whereas moderately short and short modes
change significantly.

The analyses of the correlation times of the Fourier modes show
that nanoscale objects fluctuate on the msec to $\mu$sec time range at
moderately high temperatures (400K) on Pb surfaces. Since the
atomic processes are activated, this time scale changes
dramatically with temperature.

In closing, we comment on the equilibration time of step structures
in Monte Carlo simulations. The full equilibration of these 
structures is signalled by the correlation time of the longest
wavelength mode, which can be very large (in CPU time) for the system 
sizes and temperatures studied in this paper. To do correct MC
measurements {\sl in equilibrium}, one has to pass this time; otherwise,
results for ``equilibrium quantities'' can be very misleading, as is well
known from non-equilibrium statistical mechanics. If one does not look at
correlation times of Fourier modes, very careful analysis is required 
to avoid such equilibration problems.\cite{szalma01} Recently, several papers 
have appeared concerning correlation functions,
persistence, etc., of steps much longer than ours,\cite{hgb} and sometimes even 
several of them (in studies of the interaction between them). They might well suffer
from these problems since this equilibration time scales with system size as the 
fourth power, meaning that a  system that is twice as big needs an order
of magnitude longer CPU time to be equilibrated. 
%In the multi-step case a remedy might be that
%the dynamic exponent may decrease when the steps are really close to each 
%other,\cite{pimpvillainetal93} but that is apparently (ref???????) rarely the case.

We also point out that the measurement of these fluctuations 
experimentally might be difficult
because of the above-mentioned time scale of the fluctuations. One either has
to use techniques with which snapshots of the surface can be 
taken,\cite{bartelt96} or, in direct visualization methods (like STM
measurements), the scan rate of the equipment must be faster than
the fluctuations of a given wavelength of interest. Otherwise one
measures the two ends of a wavelength at such a time separation
that they are uncorrelated, leading to difficulties of interpretation. The
``speed'' of the fluctuations can be tuned by changing the
temperature, but one also has to take into account that 
lowering the temperature decreases the size of the fluctuations,
rendering the measurement harder.

Finally, the extrapolation of our results for nano-objects to mesoscale features makes possible
comparisons of correlation times of modes of certain wavelengths, as well as of decay or of
relaxation of larger structures to their equilibrium forms. This comparison reveals 
that the relaxation of 3D structures of the same lengths are slower 
than that of the simple step or monatomic high islands, due to additional 
mechanisms and phenomena like a possible Ehrlich-Schwoebel barrier, 
very low detachment rate and perhaps elasticity that affect the 3D structure.

\section*{Acknowledgements}
Work supported by NSF MRSEC Grant DMR 00-80008, with partial
funding from NSF Grant EEC 00-85604, and the Hungarian National
Research Fund under grant No.\ OTKA D32835. We gratefully acknowledge
helpful discussions with J.R.\ Dorfman, D. Kandel, B. Koiler, H. van Beijeren, J.D.\ Weeks. Experimentalists
O. Bondarchuk, W.G. Cullen, D.B. Dougherty, and E.D. Williams provided physical motivation and insight. We thank
M.I. Haftel for calculating energy barriers for us and T.J.\ Stasevich
for computational help.

\end{document}